\documentclass[%
 longbibliography,
 reprint,onecolumn,12pt,
 amsmath,amssymb,
 aps,
]{revtex4-1}

\usepackage{graphicx}
\usepackage{dcolumn}
\usepackage{bm}
\usepackage{color} 
\usepackage{bm} 

\begin{document}

\preprint{APS/123-QED}

\title{Scattering of long water waves in a canal with rapidly varying
cross-section in the presence of a current}

\author{Semyon Churilov}
\affiliation{Institute of Solar-Terrestrial Physics of the
Siberian Branch of Russian Academy of Sciences, Irkutsk-33, PO Box
291, 664033, Russia.}
\author{Andrei Ermakov}
\affiliation{School of Agricultural, Computational and
Environmental Sciences,\\ University of Southern Queensland, QLD
4350, Australia.}%
\author{Germain Rousseaux}
\affiliation{Institut Pprime, UPR 3346, CNRS - Universit\'{e} de
Poitiers - ISAE ENSMA, 11 Boulevard Marie et Pierre Curie,
T\'{e}l\'{e}port 2, BP 30179, 86962 Futuroscope Cedex, France.}%
\author{Yury Stepanyants}
\email{Corresponding author: Yury.Stepanyants@usq.edu.au}
\affiliation{Department of Applied Mathematics, Nizhny Novgorod
State Technical University, Nizhny Novgorod, 603950, Russia and \\
School of Agricultural, Computational and Environmental Sciences,
University of Southern Queensland, QLD 4350, Australia.}%

\begin{abstract}
\vspace{2cm}

The analytical study of long wave scattering in a canal with a
rapidly varying cross-section is presented.  It is assumed that
waves propagate on a stationary current with a given flow rate.
Due to the fixed flow rate, the current speed is different in the
different sections of the canal, upstream and downstream. The
scattering coefficients (the transmission and reflection
coefficients) are calculated for all possible orientations of
incident wave with respect to the background current (downstream
and upstream propagation) and for all possible regimes of current
(subcritical, transcritical, and supercritical). It is shown that
in some cases negative energy waves can appear in the process of
waves scattering. The conditions are found when the
over-reflection and over-transmission phenomena occur. In
particular, it is shown that a spontaneous wave generation can
arise in a transcritical accelerating flow, when the background
current enhances due to the canal narrowing. This resembles a
spontaneous wave generation on the horizon of an evaporating black
hole due to the Hawking effect.
\end{abstract}

\pacs{Valid PACS appear here}
\maketitle

\section{\label{sec:level1}Introduction}

\begin{figure}[h]
\centering
\includegraphics[width=12cm]{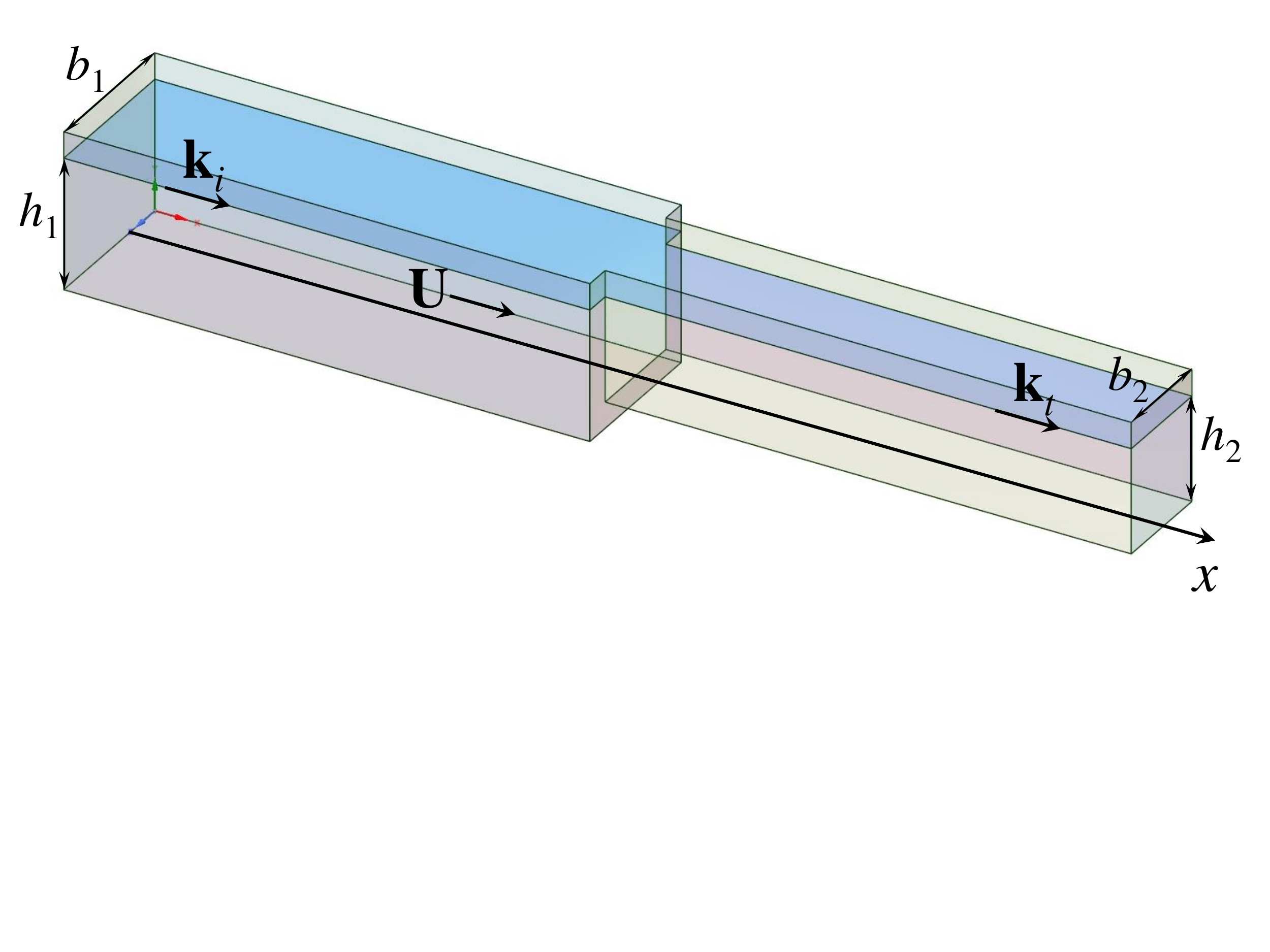}
\vspace*{-3.0cm}%
\caption{(Color online). Sketch of a canal consisting of two
sections of different rectangular cross-sections. The wave number
of incident wave is ${\bf k}_i$, and the wave number of
transmitted wave is ${\bf k}_t$ (a reflected wave is not shown).
Water flow $U$ is co-directed with the $x$-axis.}
\label{f01}%
\end{figure}

The problem of water wave transformation in a canal of a variable
cross-section is one of the classic problems of theoretical and
applied hydrodynamics. It has been studied in many books, reports,
and journal papers starting from the first edition (1879) of the
famous monograph by H. Lamb, {\em Hydrodynamics} (see the last
lifetime publication \citep{Lamb-1932}). In particular, the
coefficients of transformation of long linear waves in a canal of
a rectangular cross-section with an abrupt change of geometrical
parameters (width and depth) were presented. The transmission and
reflection coefficients were found as functions of depth ratio $X
= h_2/h_1$ and width ratio $Y = b_2/b_1$, where $h_1$ and $b_1$
are the canal depth and width at that side from which the incident
wave arrives, and $h_2$ and $b_2$ are the corresponding canal
parameters at the opposite side where the transmitted wave goes to
(see Fig. \ref{f01}). The parameters $X$ and $Y$ can be both less
than 1, and greater than 1. As explained in Ref.
\citep{Lamb-1932}, the canal cross-section can vary smoothly, but
if the wavelengths of all scattered waves are much greater than
the characteristic scale of variation of the canal cross-section,
then the canal model with the abrupt change of parameters is
valid.

The Lamb model has been further generalised for waves of arbitrary
wavelengths and applied to many practical problems. One of the
typical applications of such a model is in the problem of oceanic
wave transformation in the shelf zone; the numerous references can
be found in the books and reviews \citep{Massel-1989, Dingem-1997,
Kurkin-2015}. In such applications the canal width is assumed to
be either constant or infinitely long and only the water depth
abruptly changes.

A similar problem was studied also in application to internal
waves, but analytical results were obtained only for the
transformation coefficients of long waves in a two-layer fluid
\citep{GrimPelTal-2008}, whereas for waves of arbitrary wavelength
only the numerical results were obtained and the approximative
formulae were suggested \citep{ChurSemStep-2015}.

All aforementioned problems of wave transformation were studied
for cases when there is no background current. However, there are
many situations when there is a flow over an underwater step or in
the canals or rivers with variable cross-sections. The presence of
a current can dramatically affect the transformation coefficients
due to the specific wave-current interaction (see, e.g., Ref.
\citep{Belibassakis-2011} and references therein). The amplitudes
and energies of reflected and transmitted waves can significantly
exceed the amplitude and energy of an incident wave. Such
over-reflection and over-transmission phenomena are known in
hydrodynamics and plasma physics (see, e.g., Ref.
\citep{Jones-1968}); the wave energy in such cases can be
extracted from the mean flow. Apparently, due to complexity of
wave scattering problem in the presence of a background flow, no
results were obtained thus far even for a relatively weak flow and
small flow variation in a canal. There are, however, a number of
works devoted to wave-current interactions and, in particular,
wave scattering in spatially varying flows mainly on deep water
(see, for instance, Refs. \cite{Smith-1975, StiasDagan-79,
TrulMei-1993, Belibassakis-2011} and references therein). In Ref.
\citep{Belibassakis-2011} the authors considered the surface wave
scattering in two-dimensional geometry in $(x, y)$-plane for the
various models of underwater obstacles and currents including
vortices. In particular, they studied numerically wave passage
over an underwater step in the shoaling zone in the presence of a
current. However, the transformation coefficients were not
obtained even in the plane geometry.

Here we study the problem of long wave scattering analytically for
all possible configurations of the background flow and incident
wave (downstream and upstream propagation) in the narrowing or
widening canal (accelerating or decelerating flow) for the
subcritical, transcritical, and supercritical regimes when the
current speed is less or greater than the typical wave speed $c_0
= \sqrt{gh}$ in calm water in the corresponding canal section ($g$
is the acceleration due to gravity, and $h$ is the canal depth).
Because we consider a limiting model case of very long waves when
the variation of canal geometry is abrupt, the wave blocking
phenomenon here has a specific character of reflection. Such a
phenomenon has been studied in shallow-water limit in Ref.
\cite{Smith-1975}, but transformation coefficients were not
obtained.

Notice also that in the last decade the problem of wave-current
interaction in water with a spatially varying flow has attracted a
great deal of attention from researchers due to application to the
modelling of Hawking's radiation emitted by evaporating black
holes \citep{Unruh-1981} (see also Refs. \citep{Jacobson-1991,
Unruh-1995, Faccio-2013}). Recent experiments in a water tank
\citep{Euve-2016} have confirmed the main features of the Hawking
radiation; however many interesting and important issues are still
under investigation. In particular, it is topical to calculate the
transformation coefficients of all possible modes generated in the
process of incident mode conversion in the spatially varying flow.
Several papers have been devoted to this problem both for the
subcritical \citep{CoutWein-2016, RobMichPar-2016} and
transcritical \citep{CoutParFin-2012, Robert-2012} flows. However,
in all these papers the influence of wave dispersion was
important, whereas there is no dispersion in the problem of black
hole radiation. Our results for the dispersionless wave
transformation can shed light on the problem of mode conversion in
the relatively simple model considered in this paper.

\section{\label{sec:level2}Problem statement and dispersion relation}

\begin{figure}[h]
\centering
\includegraphics[width=15cm]{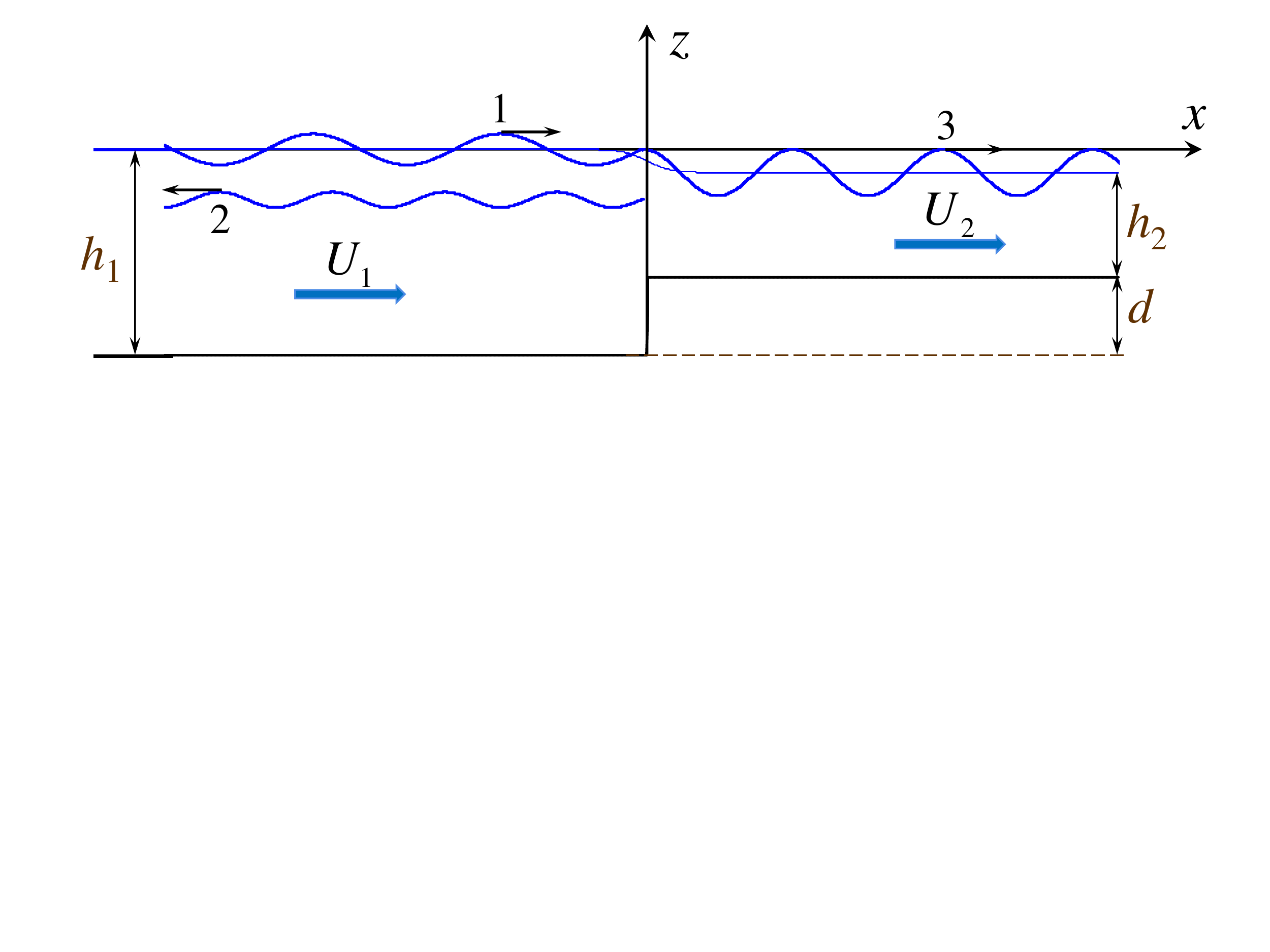}
\vspace*{-7.0cm}%
\caption{(Color online). The side view of a flow in a canal with a
variable cross-section. Wave 1 schematically represents an
incident wave, wave 2 -- a reflected wave, and wave 3 -- a
transmitted wave. The water surface slightly lowers when the
background flow increases as shown schematically by thin line.}
\label{f02}%
\end{figure}

Consider a long surface gravity wave propagating on the background
current in a canal consisting of two portions of different
cross-section each as shown in Fig. \ref{f01}. A similar problem
with a minor modification can be considered for internal waves in
two-layer fluid, but we focus here on the simplest model to gain
an insight in the complex problem of wave-current interaction. We
assume that both the canal width and depth abruptly change at the
same place, at the juncture of two canal portions. The current is
assumed to be uniform across the canal cross-section and flows
from left to right accelerating, if the canal cross-section
decreases, or decelerating, if it increases. In the presence of a
current the water surface does not remain plane even if the canal
depth is unchanged, but the width changes. According to the
Bernoulli law, when the current accelerates due to the canal
narrowing, the pressure in the water decreases and, as a result,
the level of the free surface reduces. Therefore, asymptotically,
when $x \to \infty$, the portion of canal cross-section occupied
by water is $S_2 = b_2h_2$. A similar variation in the water
surface occurs in any case when the current accelerates due to
decrease of the canal cross-section in general; this is shown
schematically in Fig. \ref{f02} (this figure is presented not in
scale, just for the sake of a vivid explanation of the wave
scattering, whereas in fact, we consider periodic waves with the
wavelengths much greater than the fluid depth).

The relationship between the water depth $h_2$, which
asymptotically onsets at the infinity, and variations of canal
width and depth at the juncture point is nontrivial. In
particular, even in the case when the canal width is unchanged,
and the canal cross-section changes only due to the presence of a
bottom step of a height $d$, the water depth $h_2$ at the infinity
is not equal to the difference $h_1 - d$ (see, e.g., Ref.
\citep{GazizMaklak-2004.}). As shown in the cited paper, variation
of a free surface due to increase of water flow is smooth even in
the case of abruptly changed depth, but in the long-wave
approximation it can be considered as abrupt. In any case, we will
parameterize the formulas for the transformation coefficients in
terms of the real depth ratio at plus and minus infinity $X =
h_2/h_1$ and canal width aspect ratio $Y = b_2/b_1$. The long-wave
approximation allows us to neglect the dispersion assuming that
the wavelength $\lambda$ of any wave participating in the
scattering is much greater than the canal depth $h$ in the
corresponding section.

In the linear approximation the main set of hydrodynamic equations
for shallow-water waves in a perfect incompressible fluid is (see,
e.g., Ref. \citep{Lamb-1932}):
\begin{eqnarray}
\frac{\partial u}{\partial t} + U\frac{\partial u}{\partial x}
&=& -g\frac{\partial \eta}{\partial x}, \label{LinEurEq} \\%
\frac{\partial \eta}{\partial t} + U\frac{\partial \eta}{\partial
x} &=& -h\frac{\partial u}{\partial x}. \label{LinMassCons} %
\end{eqnarray}
Here $u(x, t)$ is a wave induced perturbation of a horizontal
velocity, $U$ is the velocity of background flow which is equal to
$U_1$ at minus infinity and $U_2$ at plus infinity, $\eta(x, t)$
is the perturbation of a free surface due to the wave motion, and
$h$ is the canal depth which is equal to $h_1$ at minus infinity
and $h_2$ at plus infinity -- see Fig. \ref{f02}.

For the incident harmonic wave of the form $\sim
\mathrm{e}^{\mathrm{i}(\omega t - kx)}$ co-propagating with the
background flow we obtain from Eq. (\ref{LinMassCons})%
\begin{equation}
\label{Eq01}%
\left(\omega - U_1k_i\right)\eta_i = h_1k_iu_i,%
\end{equation}
where index $i$ pertains to incident wave (in what follows indices
$t$ and $r$ will be used for the transmitted and reflected waves
respectively).

Combining this with Eq. (\ref{LinEurEq}), we derive the
dispersion relation for the incident wave%
\begin{equation}
\label{DispRelIn}%
\omega = \left(U_1 + c_{01}\right)k_i,%
\end{equation}
where $c_{01} = \sqrt{gh_1}$.

Similarly for the transmitted wave we have $\left(\omega -
U_2k_t\right)\eta_t = h_2k_tu_t$ and the dispersion relation
$\omega = \left(U_2 + c_{02}\right)k_t$, where $c_{02} =
\sqrt{gh_2}$. Notice that the wave frequency remains unchanged in
the process of wave transformation in a stationary, but spatially
varying medium. Then, equating the frequencies for the incident
and transmitted waves, we obtain $k_t/k_i = \left(U_1 +
c_{01}\right)/\left(U_2 + c_{02}\right)$.

From the mass conservation for the background flow we have
$U_1h_1b_1 = U_2h_2b_2$ or $U_1/U_2 = XY$. Using this
relationship, we obtain for the wave number of the transmitted
wave
\begin{equation}
\label{WaveNumTr}%
\frac{k_t}{k_i} = XY\frac{1 + \mathrm{Fr}}{X^{3/2}Y + \mathrm{Fr}},%
\end{equation}
where $\mathrm{Fr} = U_1/c_{01}$ is the Froude number.

\begin{figure}[b]
\centering
\includegraphics[width=15cm]{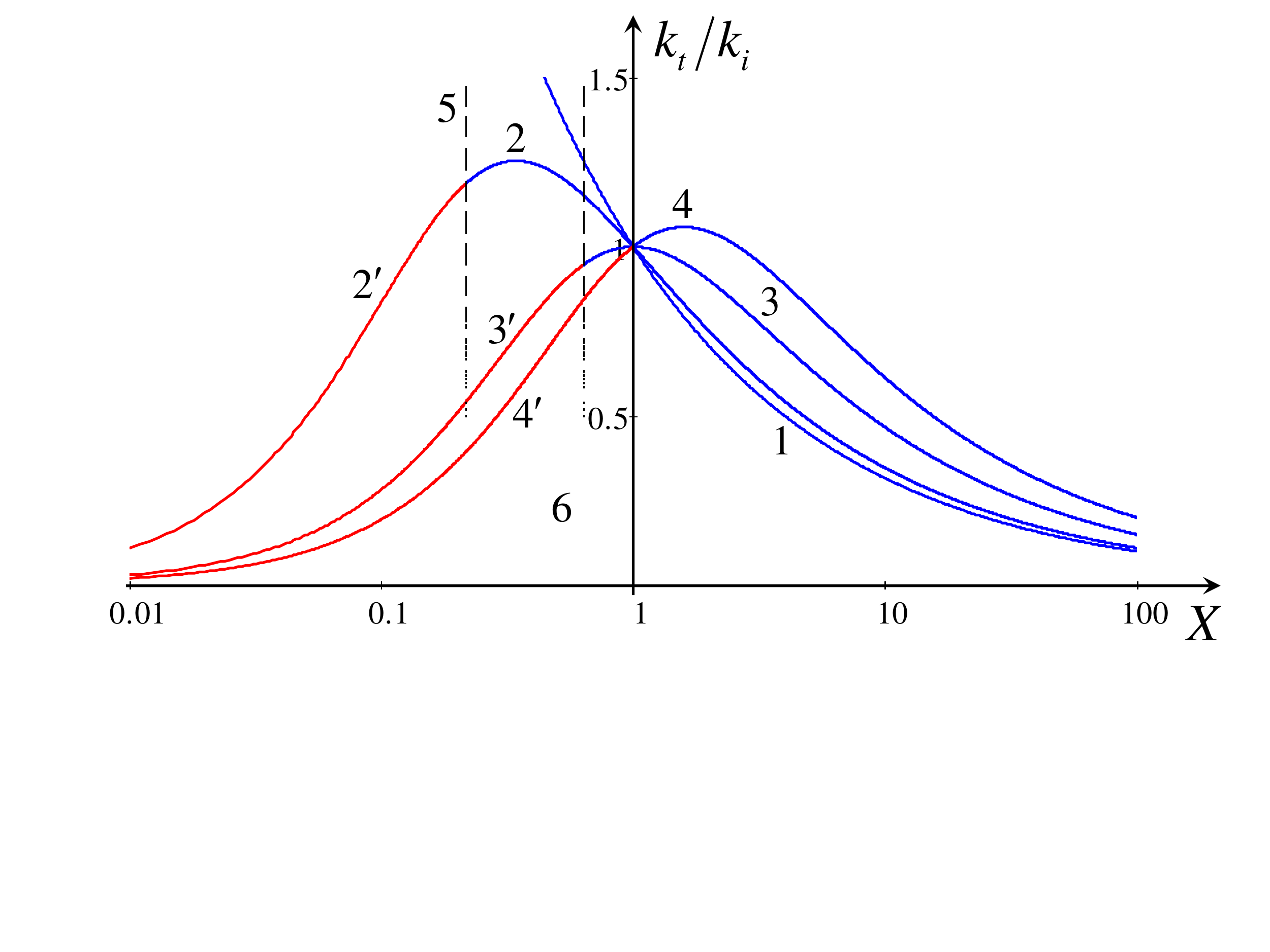}
\vspace*{-3.5cm}%
\caption{(Color online). The dependence of wave number ratio on
the depth drop $X = h_2/h_1$ for different Froude numbers and $Y =
1$. Line 1 pertains to the reference case when $\mathrm{Fr} = 0$,
lines 2 and $2'$ -- to $\mathrm{Fr} = 0.1$, lines 3 and $3'$ -- to
$\mathrm{Fr} = 0.5$, line 4 and $4'$ -- to $\mathrm{Fr} = 1$.
Dashed vertical lines 5 and 6 show the boundaries between the
subcritical and supercritical regimes in the downstream domain for
$\mathrm{Fr} = 0.1$, line 5, and $\mathrm{Fr} = 0.5$, line 6.}
\label{f03}%
\end{figure}

The relationship between the wave numbers of incident and
transmitted waves as functions of the depth drop $X$ is shown in
Fig. \ref{f03} for several values of $\mathrm{Fr}$ and $Y = 1$. As
one can see, the ratio of wave numbers $k_t/k_i$ non-monotonically
depends on $X$; it has a maximum at $X_m =
\left(2\mathrm{Fr}/Y\right)^{2/3}$. The maximum value
$\left(k_t/k_i\right)_{max} = \sqrt[3]{4Y}\left(1 +
\mathrm{Fr}\right)/\left(3\sqrt[3]{\mathrm{Fr}}\right)$ is also a
non-monotonic function of the Froude number; it has a minimum at
$\mathrm{Fr} = 0.5$ where $\left(k_t/k_i\right)_{max} =
\sqrt[3]{Y}$. In the limiting case, when there is no current
($\mathrm{Fr} = 0$), $k_t/k_i = X^{-1/2}$ independently of $Y$
(see line 1 in Fig. \ref{f03}). The current with the Froude number
$\mathrm{Fr} < 1$ remains subcritical in the downstream domain, if
$X > \left(\mathrm{Fr}/Y\right)^{2/3}$. Otherwise it becomes
supercritical. Dashed lines 5 and 6 in Fig. \ref{f03} show the
boundaries between the subcritical and supercritical regimes in
the downstream domains for two values of the Froude number,
$\mathrm{Fr} = 0.1$ and $\mathrm{Fr} = 0.5$ respectively.

For the upstream propagating reflected wave the harmonic
dependencies of free surface and velocity perturbations are
$\{\eta, u\} \sim \mathrm{e}^{\mathrm{i}(\omega t + k_rx)}$. Then
from Eq. (\ref{LinMassCons}) we obtain $\left(\omega +
U_1k_r\right)\eta_r = -h_1k_ru_r$, and combining this with Eq.
(\ref{LinEurEq}), we derive the dispersion relation for
the reflected wave with $k_r < 0$%
\begin{equation}
\label{DispRelRef}%
\omega = \left(c_{01} - U_1\right)|k_r|.%
\end{equation}

Equating the frequencies of the incident and reflected waves, we
obtain from the dispersion relations the relationship between the
wave numbers:
\begin{equation}
\label{WaveNumRef}%
\frac{|k_r|}{k_i} = \frac{1 + \mathrm{Fr}}{1 - \mathrm{Fr}}.%
\end{equation}
Notice that the ratio of wave numbers $|k_r|/k_i$ depends only on
$\mathrm{Fr}$, but does not depend on $X$ and $Y$.

\begin{figure}[h]
\centering
\includegraphics[width=15cm]{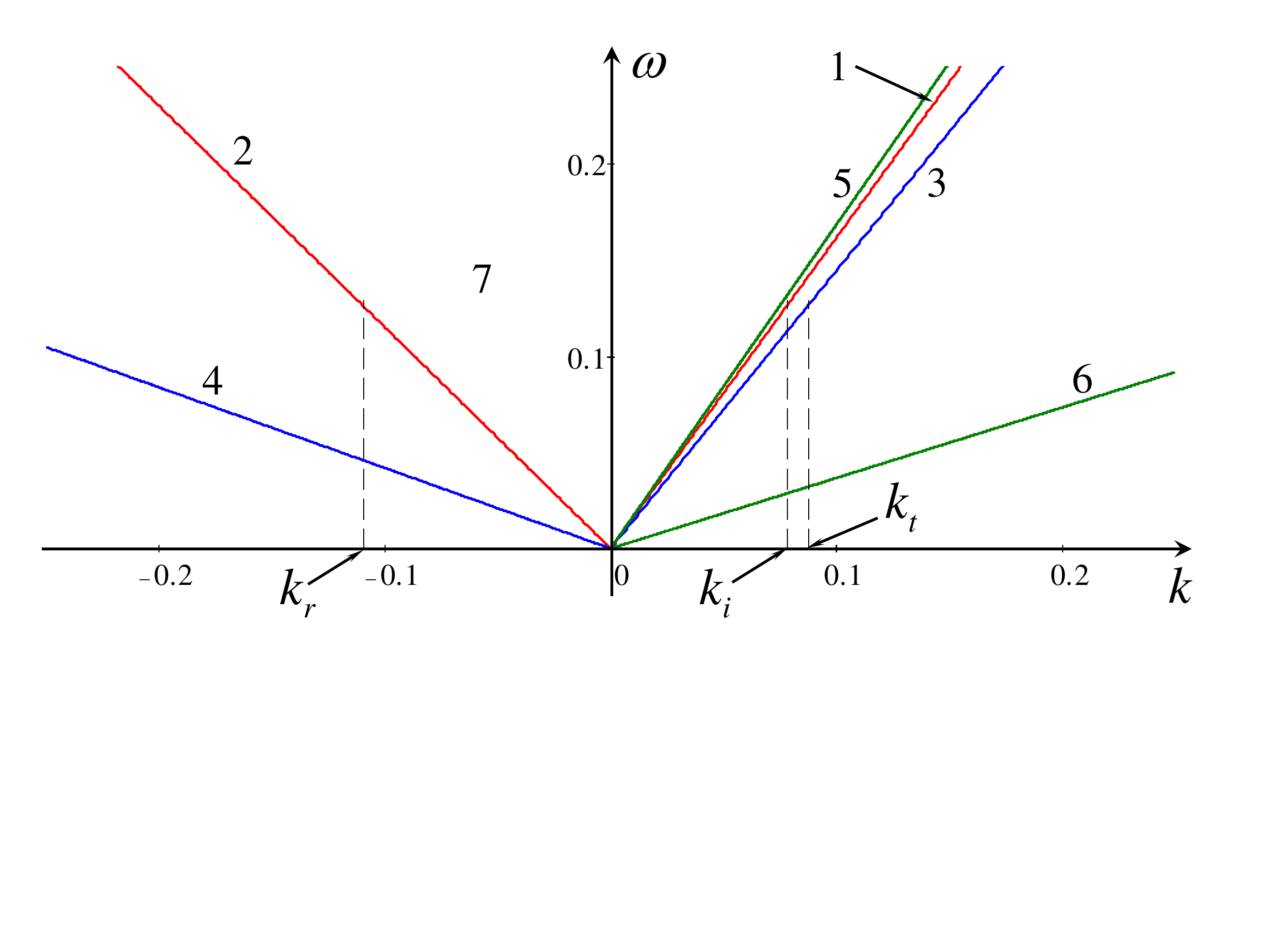}
\vspace*{-4.0cm}%
\caption{(Color online). Qualitative sketch of dispersion lines
for long surface waves on a uniform background flow in a canal.
For details see the text.}
\label{f04}%
\end{figure}

The dispersion relations for long surface waves on a constant
current are shown in Fig. \ref{f04}. Lines 1 and 2 show the
dispersion dependencies for the downstream and upstream
propagating waves, respectively, in the upstream domain, if the
background current is subcritical, i.e., when $\mathrm{Fr} < 1$.
Lines 3 and 4 show the dispersion dependencies for the downstream
and upstream propagating waves, respectively, which can
potentially exist in the downstream domain, if the background
current remains subcritical in this domain too, i.e. when
$U_2/c_{02} \equiv \mathrm{Fr}/\left(X^{3/2}Y\right) < 1$. If
there is a source generating an incident wave of frequency
$\omega$ and wave number $k_i$ at minus infinity, then after
scattering at the canal juncture the reflected wave appears in the
upstream domain with the same frequency and wave number $k_r$.
Dashed horizontal line 7 in Fig. \ref{f04} shows the given
frequency $\omega$. In the downstream domain with a subcritical
flow the incident wave generates only one transmitted wave with
the wave number $k_t$.

If the flow in one of the domains becomes faster and faster so
that $\mathrm{Fr} \to 1_-$, then the dispersion line corresponding
to the upstream propagating waves tilts to the negative portion of
horizontal axis $k$ in Fig. \ref{f04} (cf. lines 2 and 4), and its
intersection with the horizontal dashed line 7 shifts to the minus
infinity. In the case of a supercritical flow, $\mathrm{Fr}
> 1$, the dispersion line corresponding to the upstream
propagating waves is line 6 in Fig. \ref{f04}. Its intersection
with the horizontal dashed line 7 originates at the plus infinity
(as the continuation of the intersection point of line 4 with line
7 disappeared at the minus infinity) and moves to the left when
the flow velocity increases. The speeds of such waves in a calm
water are smaller than the speed of a current, therefore despite
the waves propagate counter current, the current traps them and
pulls downstream. In the immovable laboratory coordinate frame
they look like waves propagating to the right jointly with the
current. As shown in Refs. \citep{StepFabr-1989, FabrStep-1998,
MaiRusStep-2016A}, such waves possess a negative energy. This
means that the total energy of a medium when waves are excited is
less then the energy of a medium without waves. Obviously, this
can occur only in the non-equilibrium media, for example, in
hydrodynamical flows possessing kinetic energy. In the equilibrium
media, wave excitation makes the total energy greater than the
energy of the non-perturbed media (more detailed discussion of the
negative energy concept one can find in the citations presented
above and references therein). In Appendix \ref{appA} we present
the direct calculation of wave energy for the dispersionless case
considered here and show when it become negative.

With the help of dispersion relations, the links between the
perturbations of fluid velocity and free surface in the incident,
reflected and transmitted waves can be presented as
\begin{equation}
\label{Rel-u&eta}%
u_i = c_{01}\eta_i/h_1; \quad u_r = -c_{01}\eta_r/h_1; \quad u_t =
c_{02}\eta_t/h_2.%
\end{equation}

Using these relationships, we calculate in the next sections the
transformation coefficients for all possible flow regimes and
wave-current configurations.

\section{\label{sec:level3}Subcritical flow in both the upstream and downstream domains}

\subsection{\label{sec:level31}Downstream propagating incident wave}

Consider first the case when the current is co-directed with the
$x$-axis (see Fig. \ref{f02}) and the incident wave travels in the
same direction. Then, the transmitted wave is also co-directed
with the current, but the reflected wave travels against the
current. We assume that the current is subcritical in both left
domain and right domains, i.e. its speed $U_1 < c_{01}$ and $U_2 <
c_{02}$. This can be presented alternatively in terms of the
Froude number and canal specific ratios, viz $\mathrm{Fr} < 1$ and
$\mathrm{Fr} < X^{3/2}Y$.

To derive the transformation coefficients, we use the boundary
conditions at the juncture point $x = 0$. These conditions
physically imply the continuity of pressure and continuity of
horizontal mass flux induced by a surface wave. The total pressure
in the moving fluid consists of hydrostatic pressure $\rho g (h +
\eta)$ and kinetic pressure $\rho(U + u)^2/2$. The condition of
pressure continuity in the linear approximation reduces to
\begin{equation}
\label{PresCont}%
g\eta_1 + U_1u_1 = g\eta_2 + U_2u_2,%
\end{equation}
where indices 1 and 2 pertain to the left and right domains
respectively far enough from the juncture point $x = 0$. In the
left domain we have $\{\eta_1, u_1\} = \{\eta_i + \eta_r, u_i +
u_r\}$, whereas in the right domain $\{\eta_2, u_2\} = \{\eta_t,
u_t\}$.

Using the relationships between $u_{i,r,t}$ and $\eta_{i,r,t}$ as
per Eq. (\ref{Rel-u&eta}) and assuming that the incident wave has
a unit amplitude in terms of $\eta$, we obtain from Eq.
(\ref{PresCont})
\begin{equation}
\label{PresCont1}%
g\left(1 + R_{\eta}\right) + U_1\frac{c_{01}}{h_1}\left(1 -
R_{\eta}\right) = gT_{\eta} + U_2\frac{c_{02}}{h_2}T_{\eta},%
\end{equation}
where $R_{\eta}$ and $T_{\eta}$ are amplitudes of reflected and
transmitted waves respectively. In the dimensionless form this
equations reads
\begin{equation}
\label{PresCont2}%
1 + \mathrm{Fr} + \left( 1 - \mathrm{Fr}\right)R_{\eta} =
T_{\eta}\left( 1 + \frac{\mathrm{Fr}}{X^{3/2}Y}\right).
\end{equation}

The condition of mass flux continuity leads to the equation
\begin{equation}
\label{MassFluxCont}%
\rho b_1\left(h_1 + \eta_1\right)\left(U_1 + u_1\right) = \rho
b_2\left(h_2 + \eta_2\right)\left(U_2 + u_2\right).%
\end{equation}

In the linear approximation and dimensionless form this gives:
\begin{equation}
\label{MassFluxCont1}%
1 + \mathrm{Fr} - \left(1 - \mathrm{Fr}\right)R_{\eta} =
T_{\eta}\sqrt{X}Y\left(1 + \frac{\mathrm{Fr}}{X^{3/2}Y}\right).
\end{equation}

After that we derive the transformation coefficients $R_{\eta}$
and $T_{\eta}$ from Eqs. (\ref{PresCont2}) and
(\ref{MassFluxCont1}):
\begin{equation}
\label{TransCoef1}%
R_{\eta} = \frac{1 + \mathrm{Fr}}{1 - \mathrm{Fr}}\frac{1 -
\sqrt{X}Y}{1 + \sqrt{X}Y}, \quad T_{\eta} = \frac{1 +
\mathrm{Fr}}{X^{3/2}Y + \mathrm{Fr}} \frac{2X^{3/2}Y}{1 +
\sqrt{X}Y}.
\end{equation}

\begin{figure}[h]
\centering
\includegraphics[width=15cm]{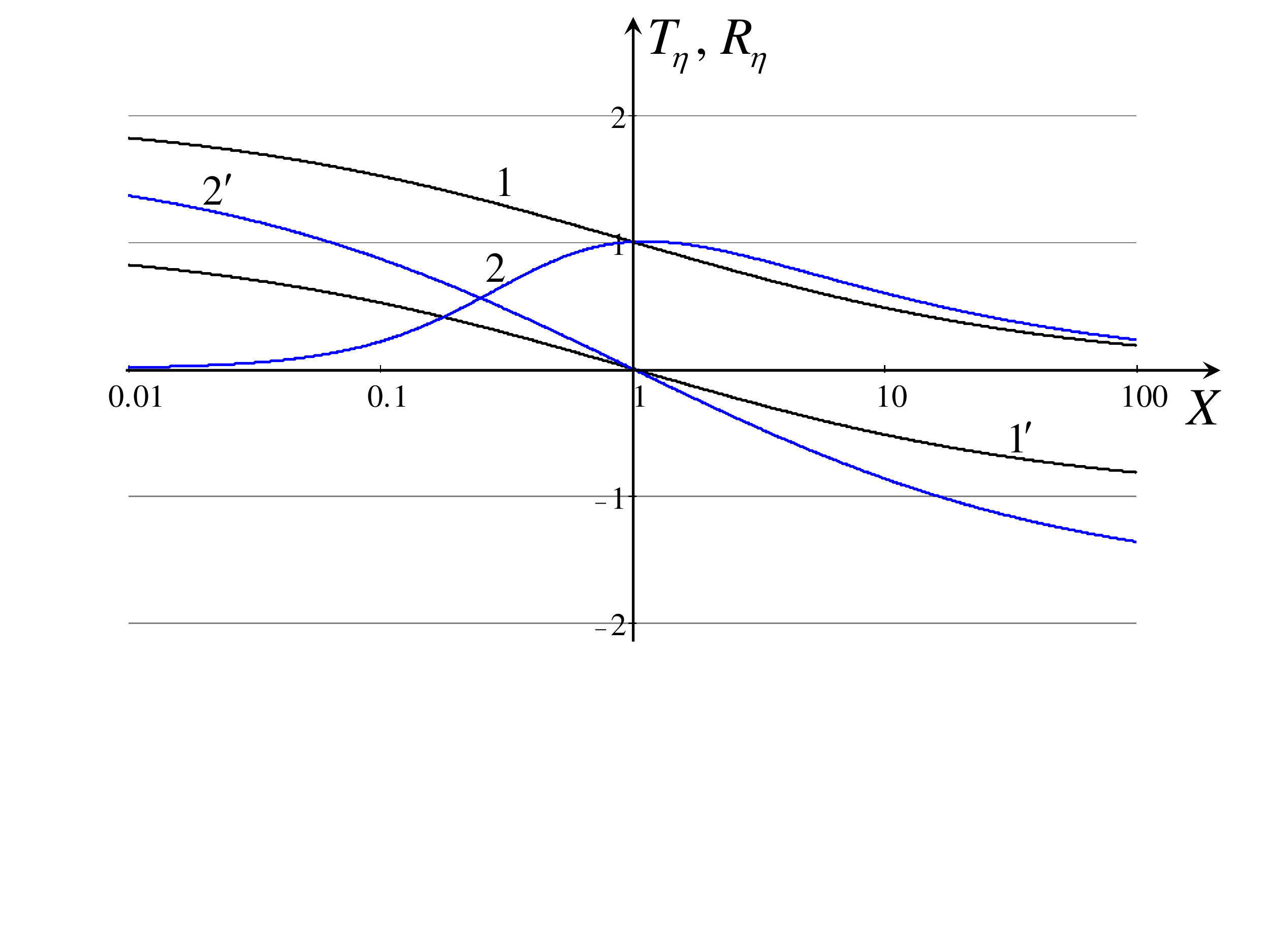}
\vspace*{-3.5cm}%
\caption{(Color online). The transformation coefficients of
surface waves on a uniform subcritical current in a canal with
flat walls, $Y = 1$, as functions of the depth drop $X$. Line 1
for $T_\eta$ and line $1'$ for $R_\eta$ pertain to the reference
case given by the Lamb formulas with $\mathrm{Fr} = 0$; lines 2
(for $T_\eta$) and $2'$ (for $R_\eta$) pertain to the flow with
$\mathrm{Fr} = 0.5$.}
\label{f05}%
\end{figure}

These formulas naturally reduce to the well-known Lamb formulas
\citep{Lamb-1932} when $\mathrm{Fr} \to 0$. Graphics of $T_{\eta}$
and $R_{\eta}$ as functions of depth drop $X$ are shown in Fig.
\ref{f05} for the particular value of Froude number $\mathrm{Fr} =
0.5$ and $Y = 1$.

As follows from the formula for $R_\eta$, the reflection
coefficient increases uniformly in absolute value, when the Froude
number increases from 0 to 1, provided that $\sqrt{X}Y \ne 1$. It
is important to notice that the reflectionless propagation can
occur in the case, when $\sqrt{X}Y = 1$, whereas neither $X$, nor
$Y$ are equal to one. The transmission coefficient in this case
$T_\eta = \left(1 + \mathrm{Fr}\right)/\left(1 +
Y^2\mathrm{Fr}\right) \ne 1$ in general, except the case when
$\mathrm{Fr} = 0$. The reflection coefficient is negative when
$\sqrt{X}Y > 1$, which means that the reflected wave is in
anti-phase with respect to the incident wave.

The dependence of $T_\eta$ on the Froude number is more
complicated and non-monotonic in $X$. However, in general
$T_{\eta} \to 0$ in two limiting cases, when $X \to 0$, then
$T_\eta \approx 2X^{3/2}Y\left(1 + 1/\mathrm{Fr}\right)$, and when
$X \to \infty$, then $T_\eta \approx 2\left(1 +
\mathrm{Fr}\right)/ \left(\sqrt{X}Y\right)$ (see Fig. \ref{f05}).

It is appropriate to mention here the nature of singularity of the
reflection coefficient $R_\eta$ and wave number $k_r$ of the
reflected wave as per Eq. (\ref{WaveNumRef}) when $\mathrm{Fr} \to
1$. In such case, the dispersion line 2 in Fig. \ref{f04}
approaches negative half-axis of $k$, and the point of
intersection of line 2 with the dashed horizontal line 7 shifts to
the minus infinity, i.e. $k_r \to -\infty$, and the wavelength of
reflected wave $\lambda_r = 2\pi/|k_r| \to 0$. Thus, we see that
when $\mathrm{Fr} \to 1$, then the amplitude of the reflected wave
$R_\eta$ infinitely increases, and its wavelength vanishes. It
will be shown below that the wave energy flux associated with the
reflected wave remains finite even when $\mathrm{Fr} = 1$.

The results obtained for the transformation coefficients are in
consistency with the wave energy flux conservation in an
inhomogeneous stationary moving fluid (see, e.g., Ref.
\citep{LongHig-1996}), $W \equiv V_gE = \mathrm{const.}$, where
$V_g \equiv d\omega/dk$ is the group speed in the moving fluid,
and $E$ is the density of wave energy. In the case of long waves
in shallow water we have $\left(V_g\right)_{1,2} =
\left(c_{0}\right)_{1,2} \pm U_{1,2}$. As shown in Appendix
\ref{appA} (see also Refs. \citep{Dysthe-2004, MaiRusStep-2016A}),
the period-averaged energy density in the long-wave limit is $E =
gA^2b\left(1 \pm \mathrm{Fr}\right)/2$, where $A$ is the amplitude
of free surface perturbation, $b$ is the canal width, sign plus
pertains to waves co-propagating with the background flow, and
sign minus -- to waves propagating against the flow. Taking into
account that the energy fluxes in the incident and transmitted
waves are directed to the right, and the energy flux in the
reflected wave is directed to the left, we obtain
\begin{equation}
\label{EnFluxCons}%
\left(1 + \mathrm{Fr}\right)^2 - \left(1 -
\mathrm{Fr}\right)^2R_{\eta}^2 = \sqrt{X}Y \left(1 +
\frac{\mathrm{Fr}}{X^{3/2}Y}\right)^2 T_{\eta}^2,
\end{equation}
where the factor $\sqrt{X}Y$ accounts for the change of the
cross-sectional area of the canal.

Substituting here the expressions for the transformation
coefficients Eq. (\ref{TransCoef1}), we confirm that Eq.
(\ref{EnFluxCons}) reduces to the identity. Notice that the second
term in the left-hand side of Eq. (\ref{EnFluxCons}), which
represents the energy flux induced by the reflected wave, remains
finite even at $\mathrm{Fr} = 1$.

The gain of energy densities in the reflected and transmitted
waves can be presented as the ratios $E_r/E_i$ and $E_t/E_i$.
Using the formulas for the transformation coefficients and
expression for the wave energy in a moving fluid (see above), we
obtain
\begin{equation}
\label{EnergyGain}%
\frac{E_r}{E_{i}} = \frac{1 + \mathrm{Fr}}{1 -
\mathrm{Fr}}\left(\frac{1 - \sqrt{X}Y}{1 + \sqrt{X}Y}\right)^2,
\quad \frac{E_t}{E_{i}} = \frac{4Y}{\left(1 +
\sqrt{X}Y\right)^2}\frac{1 + \mathrm{Fr}}{1 +
\mathrm{Fr}/X^{3/2}Y}.
\end{equation}

\begin{figure}[b]
\centering
\includegraphics[width=15cm]{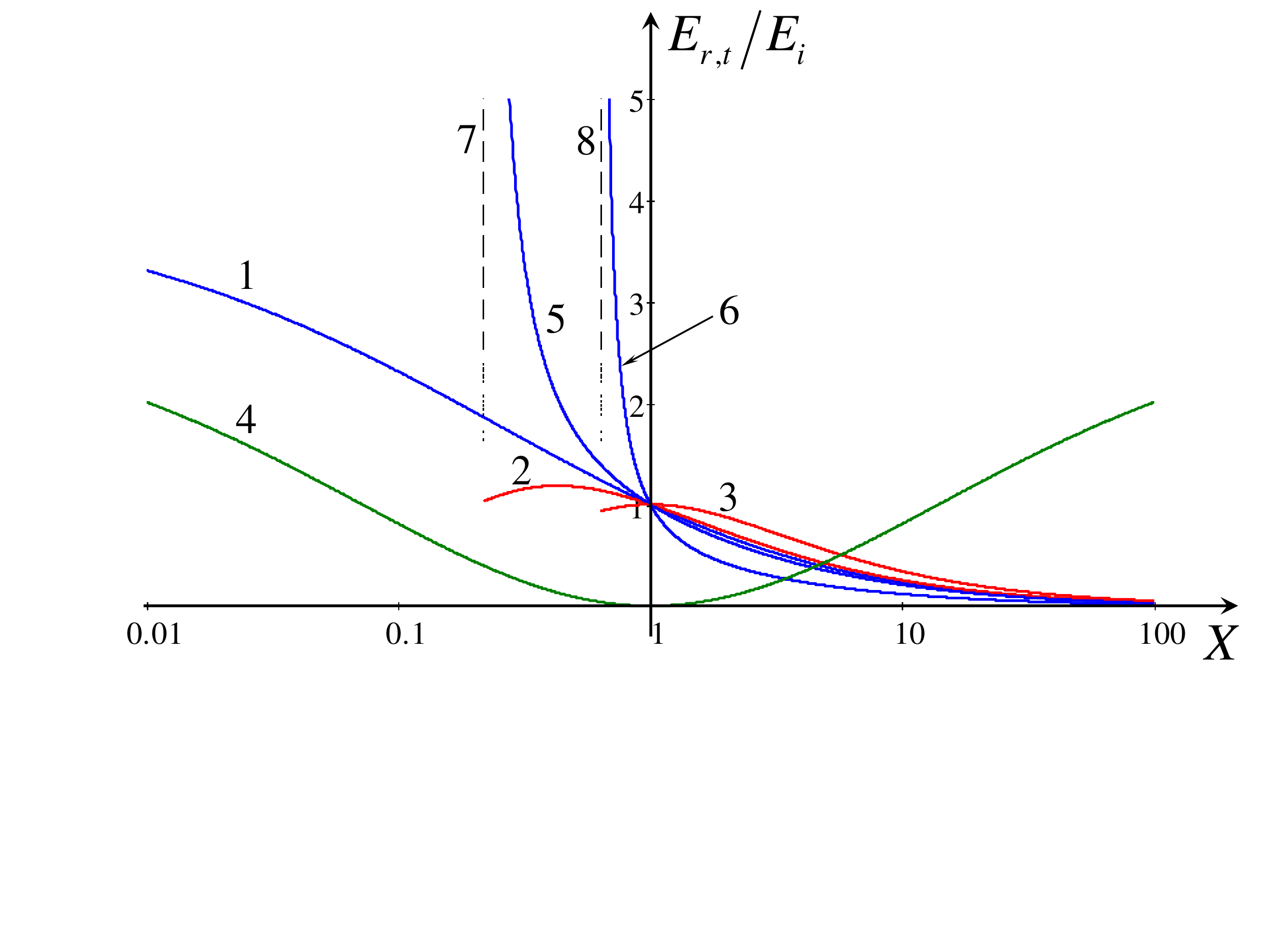}
\vspace*{-3.5cm}%
\caption{(Color online). The gain of energy density in the
transmitted wave for several Froude numbers and $Y = 1$ as
functions of the depth drop $X$. Line 1 pertains to the reference
case when $\mathrm{Fr} = 0$; lines 2 and 3 pertain to the
downstream propagating waves in the subcritical flows with
$\mathrm{Fr} = 0.1$ and 0.5 respectively; and lines 5 and 6
pertain to the upstream propagating waves in the same flows. Line
4 shows the typical dependence of energy density gain in the
upstream propagating reflected wave with $\mathrm{Fr} = 0.5$.
Lines 7 and 8 show the boundaries of subcritical regimes for
$\mathrm{Fr} = 0.1$ and 0.5 respectively.}
\label{f06}%
\end{figure}

As follows from the first of these expressions, the density of
wave energy in the reflected wave is enhanced uniformly by the
current at any Froude number ranging from 0 to 1 regardless of $X$
and $Y$, whereas the density of wave energy in the transmitted
wave can be slightly enhanced by the current only if $X^{3/2}Y >
1$; otherwise, it is less than that in the incident wave. Figure
\ref{f06} illustrates the gain of energy density in the
transmitted wave for several Froude numbers and $Y = 1$. Line 4 in
that figure shows the typical dependence of $E_r/E_i$ on $X$ for
$\mathrm{Fr} = 0.5$ and $Y = 1$. When $\mathrm{Fr} \to 1$ the gain
of wave energy in the reflected wave infinitely increases within
the framework of a linear model considered here (in reality the
nonlinear, viscous, or dispersive effects can restrict infinite
growth). In this case the typical over-reflection phenomenon
\citep{Jones-1968} occurs in the scattering of downstream
propagating wave, when the energy density in the reflected wave
becomes greater than the energy density in the incident wave. This
can occur due to the wave energy extraction from the mean flow.

\subsection{\label{sec:level32}Upstream propagating incident wave}

Consider now the case when the current is still co-directed with
the $x$ axis (see Fig. \ref{f02}) and the incident wave travels in
the opposite direction from plus infinity. Then, the transmitted
wave in the left domain propagates counter current, and the
reflected wave in the right domain is co-directed with the
current. In the dispersion diagram shown in Fig. \ref{f04} the
incident wave now corresponds to the intersection of line 2 with
the dashed horizontal line 7 (with the wave number $k_r$ replaced
by $k_i$), the reflected wave corresponds to intersection of line
1 with line 7 (with the wave number $k_i$ replaced by $k_r$), and
the transmitted wave corresponds to the intersection of line 4
with line 7 (not visible in the figure).

To derive the transformation coefficients, we use the same
boundary conditions at the juncture point $x = 0$ and after simple
manipulations similar to those presented in the previous
subsection we obtain essentially the same formulas for the wave
numbers of transmitted and reflected waves as in Eqs.
(\ref{WaveNumTr}) and (\ref{WaveNumRef}), as well as the
transformation coefficients as in Eqs. (\ref{TransCoef1}) with the
only difference that the sign of the Froude number should be
changed everywhere to the opposite, $\mathrm{Fr} \to
-\mathrm{Fr}$. However, the change of sign in the Froude number
leads to singularities in both the wave number of the transmitted
wave and the transmission coefficient. Therefore for the wave
numbers of scattered waves we obtain:
\begin{equation}
\label{WaveNumb0.22}%
\frac{k_r}{k_i} = \frac{1 - \mathrm{Fr}}{1 + \mathrm{Fr}}, \quad
\frac{k_t}{k_i} = XY\frac{1 - \mathrm{Fr}}{X^{3/2}Y -
\mathrm{Fr}}.%
\end{equation}
\begin{figure}[h]
\centering
\includegraphics[width=15cm]{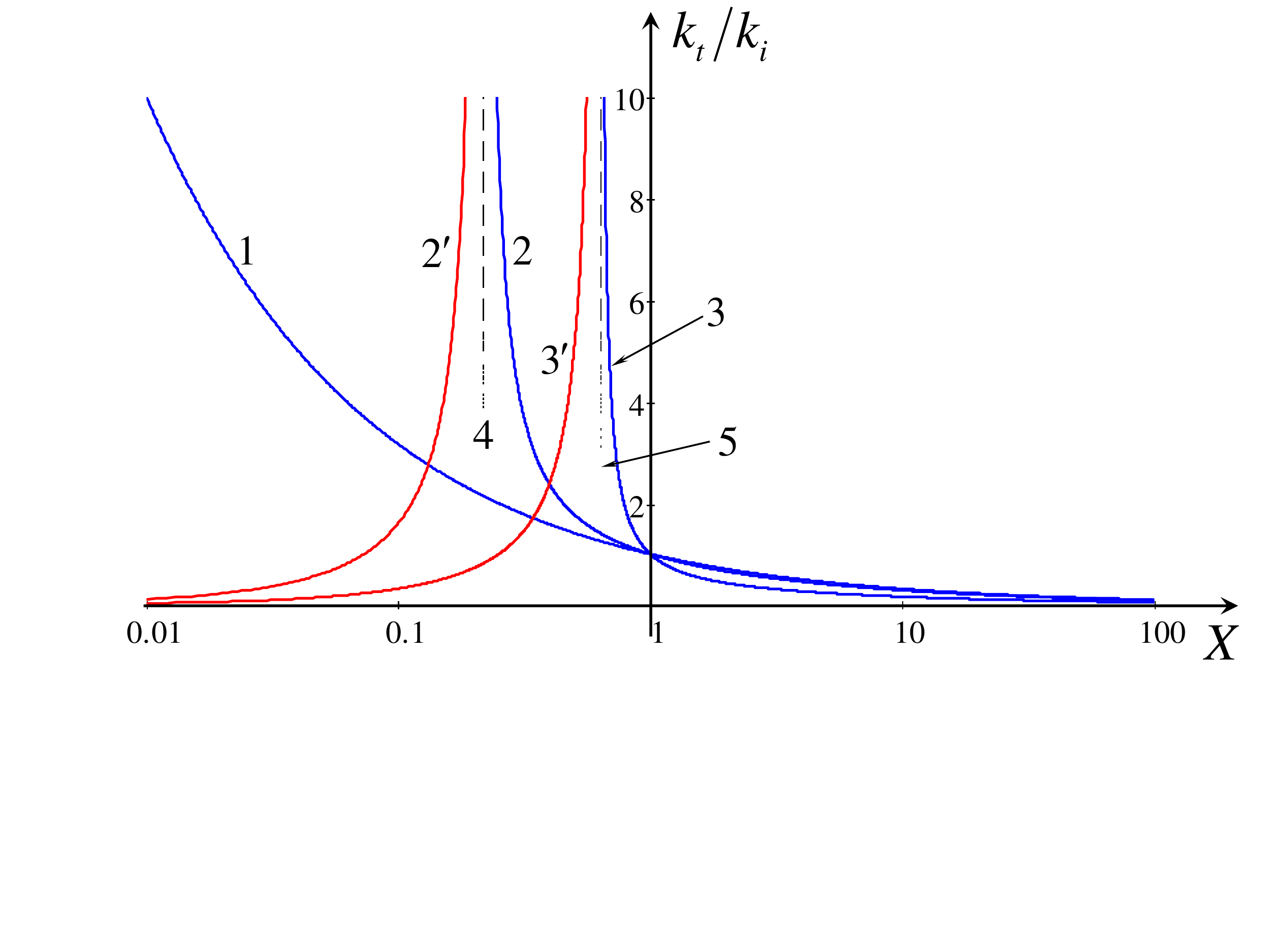}
\vspace*{-3.5cm}%
\caption{(Color online). The dependences of normalized wave
numbers of transmitted waves on the depth drop $X$ for $Y = 1$ and
several particular values of the Froude number. Line 1 pertains to
the reference case when there is no flow ($\mathrm{Fr} = 0$);
other lines pertain to the subcritical cases (line 2 --
$\mathrm{Fr} = 0.1$; line 3 -- $\mathrm{Fr} = 0.5$) and
supercritical cases (line $2'$ -- $\mathrm{Fr} = 0.1$; line $3'$
-- $\mathrm{Fr} = 0.5$). Dashed vertical lines 4 and 5 show the
boundaries between the subcritical and supercritical cases for
$\mathrm{Fr} = 0.1$ and 0.5, respectively.}
\label{f07}%
\end{figure}

In Fig. \ref{f07}, lines 1 -- 3 show the dependencies of
normalized wave numbers of transmitted waves on the depth drop $X$
for $Y = 1$ and several particular values of the Froude number.
Line 1 pertains to the reference case studied by \citet{Lamb-1932}
when there is no flow ($\mathrm{Fr} = 0$). As one can see, when
the depth drop decreases and approaches the critical value, $X \to
X_c = \left(\mathrm{Fr}/Y\right)^{2/3}$, the wave number of the
transmitted wave becomes infinitely big (and the corresponding
wavelength vanishes). This means that the current in the left
domain becomes very strong and supercritical; the transmitted wave
cannot propagate against it and the blocking phenomenon occurs
(see, e.g., Refs. \citep{BasTal-1977, MaiRusStep-2016B} and
references therein).

The transformation coefficients for this case are
\begin{equation}
\label{TransCoef0.22}%
R_{\eta} = \frac{1 - \mathrm{Fr}}{1 + \mathrm{Fr}}\frac{1 -
\sqrt{X}Y}{1 + \sqrt{X}Y}, \quad T_{\eta} = \frac{1 -
\mathrm{Fr}}{X^{3/2}Y - \mathrm{Fr}} \frac{2X^{3/2}Y}{1 +
\sqrt{X}Y}.
\end{equation}

\begin{figure}[h]
\centering
\includegraphics[width=15cm]{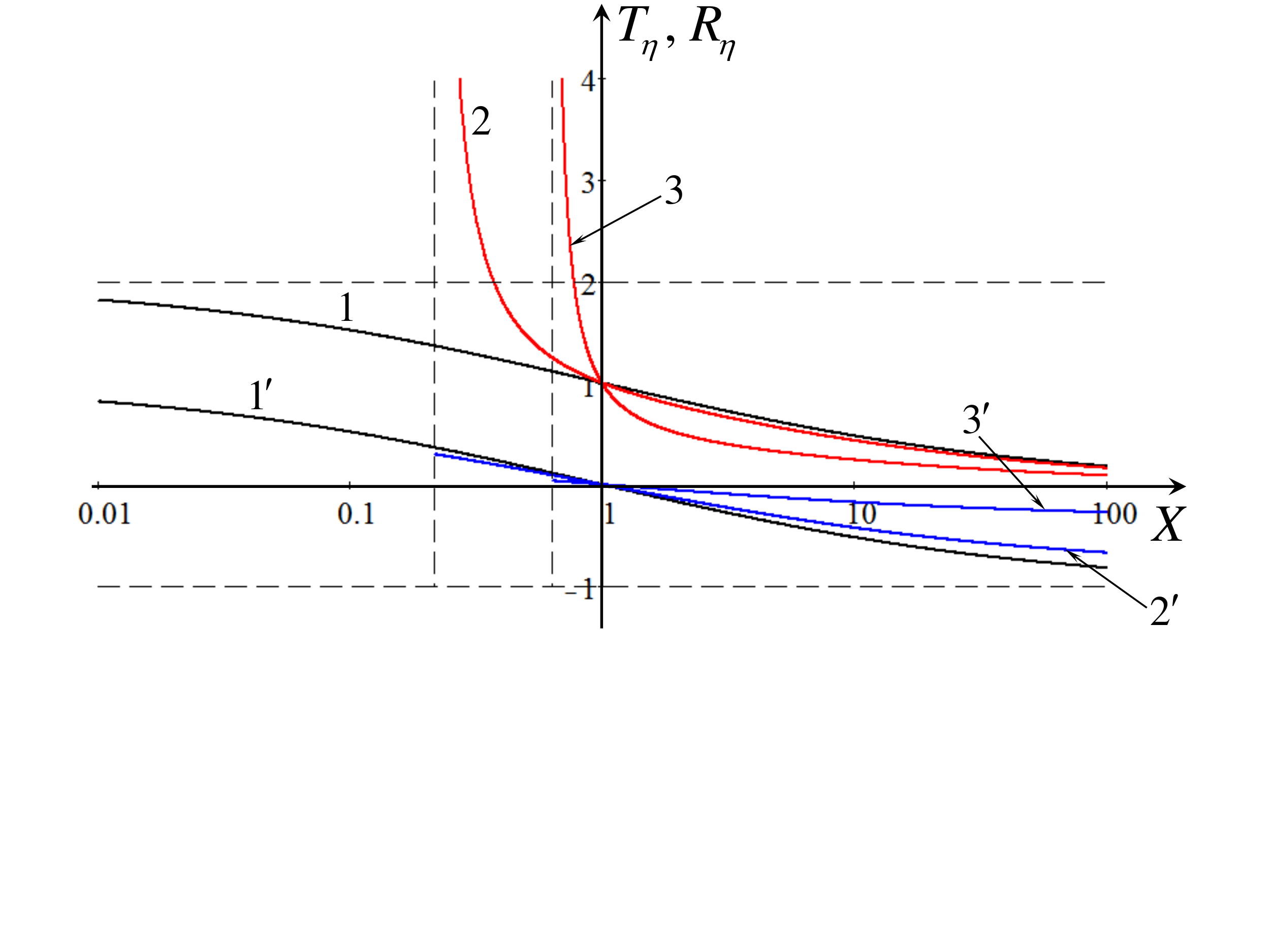}
\vspace*{-4.0cm}%
\caption{(Color online). The transformation coefficients for the
upstream propagating incident waves in a canal with flat walls, $Y
= 1$, as functions of depth drop $X$. Line 1 for $T_\eta$ and line
$1'$ for $R_\eta$ pertain to the reference case when $\mathrm{Fr}
= 0$; lines 2 (for $T_\eta$) and $2'$ (for $R_\eta$) pertain to
$\mathrm{Fr} = 0.1$, and lines 3 (for $T_\eta$) and $3'$ (for
$R_\eta$) pertain to $\mathrm{Fr} = 0.5$.}
\label{f08}%
\end{figure}

They are as shown in Fig. \ref{f08} in the domains where the
subcritical regime occurs, $X > \left(\mathrm{Fr}/Y\right)^{2/3}$
as the functions of depth drop $X$ for $Y = 1$ and two values of
the Froude number. When depth drop decreases and approaches the
critical value $X_c$, the transmission coefficient infinitely
increases, and the over-transmission phenomenon occurs. However,
it can be readily shown that the energy flux remains finite, and
the law of energy flux conservation Eq. (\ref{EnFluxCons}) with
$\mathrm{Fr} \to -\mathrm{Fr}$ holds true in this case too.

The gain of energy densities in the reflected and transmitted
waves follows from Eq. (\ref{EnergyGain}) if we replace
$\mathrm{Fr}$ by $-\mathrm{Fr}$ (see lines 4 and 5 in Fig.
\ref{f06}):
\begin{equation}
\label{EnergyGain0.22}%
\frac{E_r}{E_{i}} = \frac{1 - \mathrm{Fr}}{1 +
\mathrm{Fr}}\left(\frac{1 - \sqrt{X}Y}{1 + \sqrt{X}Y}\right)^2,
\quad \frac{E_t}{E_{i}} = \frac{4Y}{\left(1 +
\sqrt{X}Y\right)^2}\frac{1 - \mathrm{Fr}}{1 -
\mathrm{Fr}/X^{3/2}Y}.
\end{equation}

The presence of a subcritical current leads to uniform decrease of
wave energy density in the reflected wave regardless of $X$ and
$Y$. Moreover, the wave density in this wave vanishes when
$\mathrm{Fr} \to 1$. However, in the transmitted wave the density
of wave energy quickly increases when $X \to X_c$ being greater
than $X_c$ (see lines 5 and 6 in Fig. \ref{f06}). Thus, the
typical over-transmission phenomenon occurs in the scattering of
upstream propagating wave (cf. with the over-reflection phenomenon
described at the end of the previous subsection).

\section{\label{sec:level4}Subcritical flow in the upstream domain,
but supercritical in the downstream domain}

In such a case an incident wave can propagate only along the
current. In the downstream domain where the current is
supercritical no one wave can propagate against it. Therefore, we
consider here a scattering of only a downstream propagating
incident wave which arrives from minus infinity in Fig. \ref{f01}.
We assume that the Froude number and geometric parameters of a
canal are such that $X^{3/2}Y < \mathrm{Fr} < 1$.

In the upstream domain two waves of frequency $\omega$ can
propagate in the subcritical flow. One of them is an incident wave
with the unit amplitude and wave number $k_i = \omega/(c_{01} +
U_1)$ and another one is the reflected wave with the amplitude
$R_\eta$ and wave number $k_r = \omega/(c_{01} - U_1)$. In the
downstream domain two waves can exist too. One of them is the
transmitted wave of positive energy with the amplitude $T_p$ and
wave number $k_{t1} = \omega/(U_2 + c_{02})$ and another one is
the transmitted wave of negative energy (see the Appendix) with
the amplitude $T_n$ and wave number $k_{t2} = \omega/(U_2 -
c_{02})$.

The relationships between the wave numbers of scattered waves
follows from the frequency conservation. For the transmitted wave
of positive energy and reflected wave we obtain the same formulas
as in Eqs. (\ref{WaveNumTr}) and (\ref{WaveNumRef}), whereas for
the transmitted wave of negative energy we obtain
\begin{equation}
\label{WaveNumTr2}%
\frac{k_{t2}}{k_i} = XY\frac{\mathrm{Fr} + 1}{\mathrm{Fr} - X^{3/2}Y}.%
\end{equation}

As follows from this formula, the wave number $k_{t2}$ infinitely
increases when $X \to X_c$ being less than $X_c$. The dependencies
of $k_{t1}/{k_i}$ are shown in Fig. \ref{f03} by lines $2'$, $3'$,
and $4'$ for $\mathrm{Fr} = 0.1, \; 0.5$, and 1, respectively,
whereas the dependencies of $k_{t2}/{k_i}$ are shown in Fig.
\ref{f07} by lines $2'$ and $3'$ for $\mathrm{Fr} = 0.1$ and 0.5
respectively.

To find the transformation coefficients we use the same boundary
conditions as in Eqs. (\ref{PresCont1}) and (\ref{MassFluxCont}),
but now they provide the following set of equations:
\begin{eqnarray}
1 + \mathrm{Fr} + \left( 1 - \mathrm{Fr}\right)R_{\eta} &=&
T_{p}\left( 1 + \frac{\mathrm{Fr}}{X^{3/2}Y}\right) + T_{n}\left(1
- \frac{\mathrm{Fr}}{X^{3/2}Y}\right), \label{PCond1} \\%
1 + \mathrm{Fr} - \left( 1 - \mathrm{Fr}\right)R_{\eta} &=&
\sqrt{X}Y\left[T_{p}\left( 1 + \frac{\mathrm{Fr}}{X^{3/2}Y}\right)
- T_{n}\left(1 - \frac{\mathrm{Fr}}{X^{3/2}Y}\right)\right].
\label{PCond2}%
\end{eqnarray}

This set relates three unknown quantities, $R_\eta$, $T_p$, and
$T_n$. We can express, for example, amplitudes of transmitted
waves $T_p$ and $T_n$ in terms of unit amplitude of incident wave
and amplitude of reflected wave $R_\eta$:
\begin{eqnarray}
T_p &=& \frac{X}{2\left(X^{3/2}Y +
\mathrm{Fr}\right)}\left[\left(1 +
\mathrm{Fr}\right)\left(\sqrt{X}Y + 1\right) + \left(1 -
\mathrm{Fr}\right)\left(\sqrt{X}Y - 1\right)R_\eta\right],
\label{Sub-superT1} \\%
T_n &=& \frac{X}{2\left(X^{3/2}Y -
\mathrm{Fr}\right)}\left[\left(1 +
\mathrm{Fr}\right)\left(\sqrt{X}Y - 1\right) + \left(1 -
\mathrm{Fr}\right)\left(\sqrt{X}Y + 1\right)R_\eta\right],
\label{Sub-superT2}%
\end{eqnarray}
whereas the reflection coefficient $R_\eta$ remains unknown.

It can be noticed a particular case when the background flow
could, probably, spontaneously generate waves to the both sides of
a juncture where the background flow abruptly changes from the
subcritical to supercritical value. Bearing in mind that the
transformation coefficients are normalized on the amplitude of an
incident wave, $R_\eta \equiv A_r/A_i$, $T_p \equiv A_p/A_i$, $T_n
\equiv A_n/A_i$, and considering a limit when $A_i \to 0$, we
obtain from Eqs.~(\ref{Sub-superT1}) and (\ref{Sub-superT2}):
\begin{equation}
\label{SpontGen}%
\frac{A_r}{A_p} = \frac{2}{X\left(1 -
\mathrm{Fr}\right)}\frac{X^{3/2} + \mathrm{Fr}}{\sqrt{X}Y - 1},
\quad \frac{A_n}{A_p} = \frac{\sqrt{X}Y + 1}{\sqrt{X}Y -
1}\frac{X^{3/2} + \mathrm{Fr}}{X^{3/2} - \mathrm{Fr}}.%
\end{equation}

The conservation of wave energy flux in general is
\begin{equation}
\label{EnFluxCons0}%
\left(1 + \mathrm{Fr}\right)^2 - \left(1 -
\mathrm{Fr}\right)^2R_\eta^2 = \frac{1}{X^{5/2}Y}\left[
\left(X^{3/2}Y + \mathrm{Fr}\right)^2T_p^2 - \left(X^{3/2}Y -
\mathrm{Fr}\right)^2T_n^2\right].
\end{equation}

After substitution here of the transmission coefficients Eqs.
(\ref{Sub-superT1}) and (\ref{Sub-superT2}) we obtain the identity
regardless of $R_\eta$. In the case of spontaneous wave generation
when there is no incident wave, Eq.~(\ref{EnFluxCons0}) turns to
the identity too after its re-normalization and substitution of
Eqs. (\ref{SpontGen}). This resembles a spontaneous wave
generation due to Hawking's effect \citep{Unruh-1981, Unruh-1995,
Faccio-2013}) at the horizon of an evaporating black hole, when a
positive energy wave propagates towards our space (the upstream
propagating wave $A_r$ in our case), whereas a negative energy
wave together with a positive energy wave propagates towards the
black hole (the downstream propagating waves $A_n$ and $A_p$).

Thus, within the model with an abrupt change of canal
cross-section the complete solution for the wave scattering cannot
be obtained in general. One needs to discard from the
approximation when the current speed abruptly increases at the
juncture and consider a smooth current transition from one value
$U_1$ to another one $U_2$ (this problem was recently studied in
Ref. \citep{ChurErStep-2017}).

\section{\label{sec:level5}Supercritical flow in both the upstream
and downstream domains}

Now let us consider a wave scattering in the case when the flow is
supercritical both in upstream and downstream domain, $U_1 >
c_{01}$ and $U_2 > c_{02}$. In terms of the Froude number we have
$\mathrm{Fr} > 1$ and $\mathrm{Fr} > X^{3/2}Y$. It is clear that
in such a situation, similar to the previous subsection, only a
downstream propagating incident wave can be considered.

In the upstream supercritical flow there is no reflected wave. In
the dispersion diagram of Fig. \ref{f04} the downstream
propagating incident wave of frequency $\omega$ can be either the
wave on the intersection of line 5 with the dashed horizontal
line, or on the intersection of line 6 with the dashed horizontal
line (the intersection point is off the figure), or even both. The
former wave is the wave of positive energy and has the wave number
$k_{i1} = \omega/(U_1 + c_{01})$, whereas the latter is the wave
of negative energy (see the Appendix) and has the wave number
$k_{i2} = \omega/(U_1 - c_{01})$.

In the downstream domain where we assume that the flow is
supercritical too, two waves appear as the result of scattering of
incident waves. As in the upstream domain, one of the transmitted
waves has positive energy and the wave number $k_{t1} =
\omega/(U_2 + c_{02})$, and the other has negative energy and the
wave number $k_{t2} = \omega/(U_2 - c_{02})$.

Let us assume that there is a wavemaker at minus infinity that
generates a sinusoidal surface perturbation of frequency $\omega$.
Then, two waves of positive and negative energies with the
amplitudes $A_p$ and $A_n$, respectively, can jointly propagate.
In the process of wave scattering at the canal juncture two
transmitted waves with opposite energies will appear with the
amplitudes $T_p$ and $T_n$. Their amplitudes can be found from the
boundary conditions Eqs. (\ref{PresCont1}) and
(\ref{MassFluxCont}). Then, after simple manipulations similar to
those in Secs. \ref{sec:level3} and \ref{sec:level4} we obtain:
\begin{eqnarray}
T_p &=& \frac{X}{2\left(X^{3/2}Y +
\mathrm{Fr}\right)}\left[\left(\mathrm{Fr +
1}\right)\left(\sqrt{X}Y + 1\right)A_p - \left(\mathrm{Fr} -
1\right)\left(\sqrt{X}Y - 1\right)A_n\right],
\label{Super-superT1} \\%
T_n &=& \frac{X}{2\left(X^{3/2}Y -
\mathrm{Fr}\right)}\left[\left(\mathrm{Fr +
1}\right)\left(\sqrt{X}Y - 1\right)A_p - \left(\mathrm{Fr} -
1\right)\left(\sqrt{X}Y + 1\right)A_n\right].
\label{Super-superT2}%
\end{eqnarray}

At certain relationships between the amplitudes $A_p$ and $A_n$ it
may happen that there is only one  transmitted wave, either of
positive energy ($T_n = 0$), when
\begin{equation}
\label{ZeroNEWave}%
A_n = A_p\frac{\mathrm{Fr} + 1}{\mathrm{Fr} - 1}\frac{\sqrt{X}Y -
1}{\sqrt{X}Y + 1},
\end{equation}
or of negative energy ($T_p = 0$), when
\begin{equation}
\label{ZeroPEWave}%
A_n = A_p\frac{\mathrm{Fr} + 1}{\mathrm{Fr} - 1}\frac{\sqrt{X}Y +
1}{\sqrt{X}Y - 1}.
\end{equation}

From the law of wave energy flux conservation we obtain
\begin{equation}
\label{EnFluxCons1}%
\left(\mathrm{Fr} + 1\right)^2A_p^2 - \left(\mathrm{Fr} -
1\right)^2A_n^2 = \sqrt{X}Y\left[
\left(\frac{\mathrm{Fr}}{X^{3/2}Y} + 1\right)^2T_p^2 -
\left(\frac{\mathrm{Fr}}{X^{3/2}Y} - 1\right)^2T_n^2\right].
\end{equation}

Substituting here the expressions for $T_p$ and $T_n$ as per Eqs.
(\ref{Super-superT1}) and (\ref{Super-superT2}), we see that Eq.
(\ref{EnFluxCons1}) becomes an identity regardless of amplitudes
of incoming waves $A_p$ and $A_n$, including the cases when they
are related by Eqs.~(\ref{ZeroNEWave}) or (\ref{ZeroPEWave}). In
the particular cases one of the incident waves can be suppressed,
ether the wave of negative energy or wave of positive energy. In
the former case we set $A_n = 0$ and $A_p = 1$, and in the latter
case we set $A_p = 0$ and $A_n = 1$.

When there is only one incident wave of {\it positive energy} with
the amplitude $A_p = 1$ and there is no wave of negative energy
($A_n = 0$), then the transmission coefficients Eqs.
(\ref{Super-superT1}) and (\ref{Super-superT2}) reduce to
\begin{equation}
\label{Super-superT3}%
T_p = \frac{X}{2}\frac{\mathrm{Fr} + 1}{\mathrm{Fr} +
X^{3/2}Y}\left(1 + \sqrt{X}Y\right), \quad T_n =
\frac{X}{2}\frac{\mathrm{Fr} + 1}{\mathrm{Fr} - X^{3/2}Y}\left(1 -
\sqrt{X}Y\right).
\end{equation}

Recall that these formulas are valid for supercritical flows when
$\mathrm{Fr} > 1$ and $\mathrm{Fr} > X^{3/2}Y$. In the limiting
case when $X \to 0$ and $Y = \mathrm{const.}$, we obtain
\begin{equation}
\label{Super-superT4}%
T_p \approx T_n \approx X\frac{\mathrm{Fr} + 1}{\mathrm{2Fr}}.
\end{equation}

\begin{figure}[h]
\centering
\includegraphics[width=16cm]{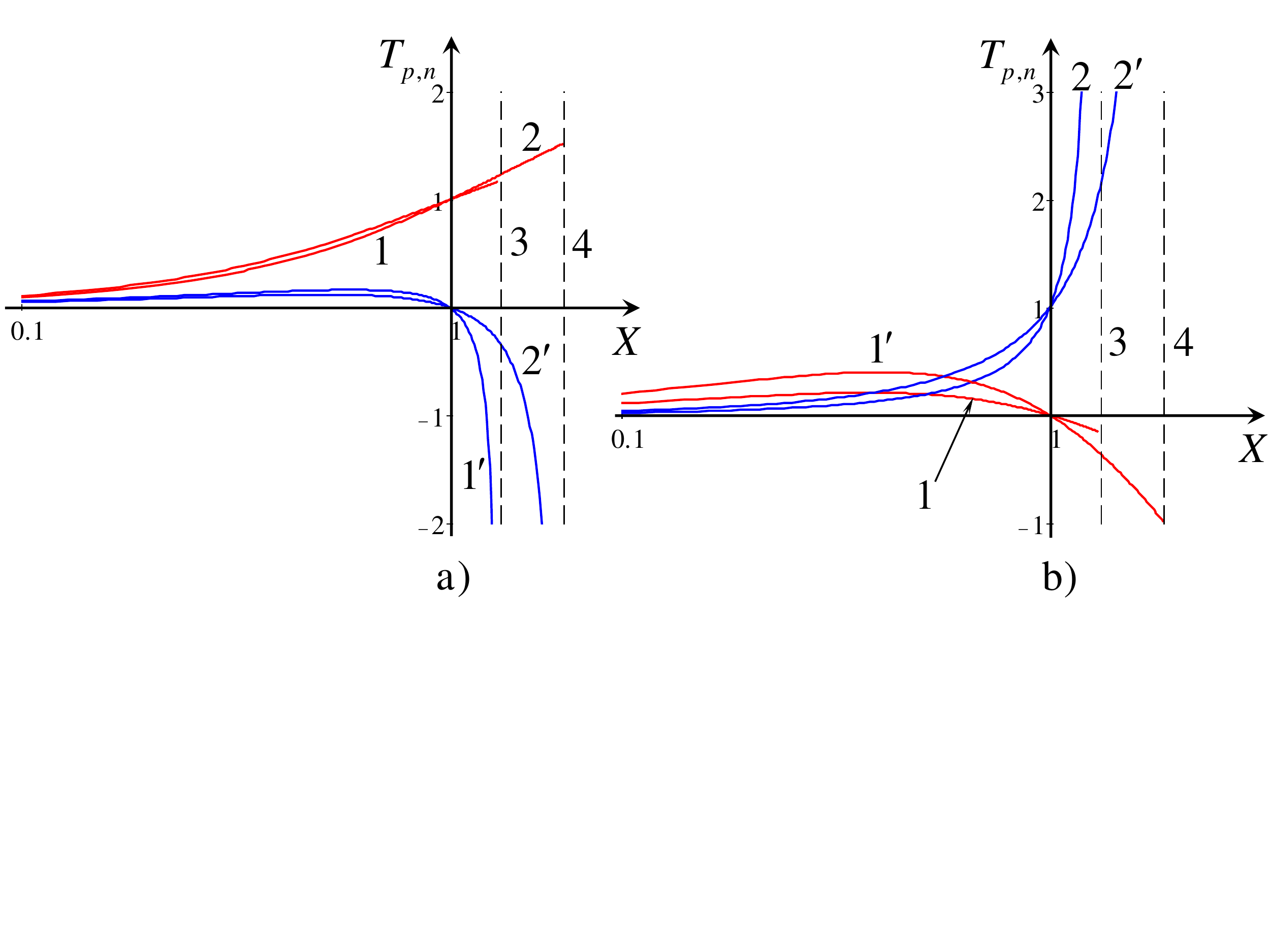}
\vspace*{-4.5cm}%
\caption{(Color online). The transmission coefficients for the
downstream propagating incident waves of positive energy (frame a)
and negative energy (frame b) in a canal with the flat walls, $Y =
1$, as functions of the depth drop $X$. Line 1 for $T_p$ and line
$1'$ for $T_n$ pertain to $\mathrm{Fr} = 1.5$, and lines 2 (for
$T_p$) and $2'$ (for $T_n$) pertain to $\mathrm{Fr} = 2.5$. Data
for lines 1 and 2 in frame (b) were multiplied by a factor of ten
to make the graphics clearly visible.}
\label{f09}%
\end{figure}

In another limiting case when $X^{3/2}Y \to \mathrm{Fr}$ the
transmission coefficient for the positive energy wave remains
constant, whereas the transmission coefficient for the negative
energy wave within the framework of linear theory goes to plus or
minus infinity depending on the value of $Y$. Figure \ref{f09}(a)
illustrates the transmission coefficients $T_p$ and $T_n$ as
functions of $X$ for $Y = 1$ and two particular values of the
Froude number.

When there is only one incident wave of {\it negative energy} with
the amplitude $A_n = 1$ and there is no wave of positive energy
($A_p = 0$), then the transmission coefficients Eqs.
(\ref{Super-superT1}) and (\ref{Super-superT2}) reduce to
\begin{equation}
\label{Super-superT5}%
T_p = \frac{X}{2}\frac{\mathrm{Fr} - 1}{\mathrm{Fr} +
X^{3/2}Y}\left(1 - \sqrt{X}Y\right), \quad T_n =
\frac{X}{2}\frac{\mathrm{Fr} - 1}{\mathrm{Fr} - X^{3/2}Y}\left(1 +
\sqrt{X}Y\right).
\end{equation}

In the limiting case when $X \to 0$, and $Y = \mathrm{const.}$, we
obtain
\begin{equation}
\label{Super-superT6}%
T_p \approx T_n \approx X\frac{\mathrm{Fr} - 1}{2\mathrm{Fr}}.
\end{equation}

In another limiting case when $X^{3/2}Y \to \mathrm{Fr}$, the
transmission coefficient for the positive energy wave remains
finite, whereas, the transmission coefficient for the negative
energy wave within the framework of linear theory goes to plus
infinity. Figure \ref{f09}(b) shows the transmission coefficients
$T_p$ and $T_n$ as functions of $X$ for $Y = 1$ for two particular
values of the Froude number.

\section{\label{sec:level6}Supercritical flow in the upstream and
subcritical in the downstream domain}

Let us consider, at last, the case when the flow is supercritical
in the upstream domain, where $U_1 > c_{01}$, but due to canal
widening becomes subcritical in the downstream domain, where $U_2
< c_{02}$. Thus, the flow is decelerating and in terms of the
Froude number we have $1 < \mathrm{Fr} < X^{3/2}Y$. Assume first
that the incident wave propagates downstream.

\subsection{\label{sec:level61}Downstream propagating incident wave}

As was mentioned in the previous section, two waves with the
amplitudes $A_p$ and $A_n$ can propagate simultaneously from minus
infinity, if they are generated by the same wavemaker with the
frequency $\omega$. In the downstream domain potentially two waves
of positive energy can exist, but only one of them propagating
downstream can appear as the transmitted wave with the amplitude
$T_\eta$ as the result of wave scattering at the juncture.

The amplitudes of scattered waves can be found from the boundary
conditions Eqs. (\ref{PresCont1}) and (\ref{MassFluxCont}). This
gives, after simple manipulations:
\begin{eqnarray}
\left(1 + \mathrm{Fr}\right)A_p + \left(1 - \mathrm{Fr}\right)A_n
&=& T_{\eta}\left(1 + \frac{\mathrm{Fr}}{X^{3/2}Y}\right), \label{PCond51} \\%
\left(1 + \mathrm{Fr}\right)A_p - \left(1 - \mathrm{Fr}\right)A_n
&=& \sqrt{X}YT_{\eta}\left( 1 +
\frac{\mathrm{Fr}}{X^{3/2}Y}\right). \label{PCond52}%
\end{eqnarray}

This set of equations provides a unique solution for the
transmission coefficient $T_\eta$ only in the case when the
amplitudes of incoming waves are related:
\begin{equation}
\label{RelApAn}%
A_n = \frac{1 + \mathrm{Fr}}{1 - \mathrm{Fr}}\frac{1 -
\sqrt{X}Y}{1 + \sqrt{X}Y}A_p, \quad T_\eta = \frac{1 +
\mathrm{Fr}}{X^{3/2}Y + \mathrm{Fr}}\frac{2X^{3/2}Y}{1 +
\sqrt{X}Y}A_p.
\end{equation}

If one of the incident waves is absent ($A_n = 0$ or $A_p = 0$) or
amplitudes of incoming waves are not related by Eq.
(\ref{RelApAn}), then the set of Eqs. (\ref{PCond51}) and
(\ref{PCond52}) is inconsistent. In such cases the problem of wave
scattering in the canal does not have a solution within the
framework of a model with a sharp change of the cross-section.

If the amplitudes of incident waves $A_p$ and $A_n$ are related by
Eq. (\ref{RelApAn}), then the conservation of wave energy flux
holds and takes the form
\begin{equation}
\label{EnFluxCons51}%
\left(\mathrm{Fr} + 1\right)^2A_p^2 - \left(\mathrm{Fr} -
1\right)^2A_n^2 = \sqrt{X}Y \left(\frac{\mathrm{Fr}}{X^{3/2}Y} +
1\right)^2T_\eta^2.
\end{equation}

Substituting here $A_n$ and $T_\eta$ from Eq. (\ref{RelApAn}), we
see that it becomes just the identity.

\subsection{\label{sec:level62}Upstream propagating incident wave}

For the incident wave arriving from the plus infinity and
propagating upstream in the subcritical domain of the flow, the
problem of wave scattering within the model with a sharp change of
a current is undefined. The incoming wave cannot penetrate from
the domain with a subcritical flow into the domain with a
supercritical flow, therefore one can say that formally the
reflection coefficient in this case $R_\eta = 1$, and the
transmission coefficients $T_\eta = 0$. However such a problem
should be considered within a more complicated model with a smooth
transcritical flow; this has been done in Ref.
\citep{ChurErStep-2017}.

\section{\label{sec:level7}Conclusion}

In this paper within the linear approximation we have studied a
scattering of long surface waves at the canal juncture when its
width and depth abruptly change at a certain place. We have
calculated the transformation coefficients for the reflected and
transmitted waves in the presence of a background flow whose speed
changes from $U_1$ to $U_2$ in accordance with the mass flux
conservation. The calculated coefficients represent the
effectiveness of the conversion of the incident wave into the
other wave modes -- reflected and transmitted of either positive
or negative energy. Our consideration generalizes the classical
problem studied by Ref. \cite{Lamb-1932} when the background flow
is absent. It was assumed that the characteristic scale of current
variation in space is much less than the wavelengths of scattered
waves. Such a simplified model allows one to gain insight into the
complex problem of wave-current interaction and find the
conditions for the over-reflection and over-transmission of water
waves. We have analyzed all possible orientations of the incident
wave with respect to flow and studied all possible regimes of
water flow (subcritical, supercritical, and transcritical).

In the study of the subcritical and supercritical flows (see Secs.
\ref{sec:level3} and \ref{sec:level5}) we have succeeded in
calculating the transmission and reflection coefficients in the
explicit forms as functions of the depth drop $X = h_2/h_1$,
specific width ratio $Y = b_2/b_1$, and Froude number
$\mathrm{Fr}$. Based on these, the conditions for the
over-reflection and over-transmission have been found in terms of
the relationships between the Froude number and canal geometric
parameters $X$ and $Y$. It appears that it is not possible to do
the same for the transcritical flows, at least within the
framework of the simplified model considered in this paper (see
Secs. \ref{sec:level4} and \ref{sec:level6}). The reason for that
is in the critical point where $\mathrm{Fr} = 1$ which appears in
the smooth transient domain between two portions of a canal with
the different cross-sections. The transition through the critical
point is a rather complex problem which was recently studied on
the basis of a model with a continuously varying flow speed in a
duct of smoothly varying width \citep{ChurErStep-2017}. The
summary of results obtained is presented in
Table I.\\

Table I. The summary of considered cases. A cocurrent propagating
incident waves is denoted by $k_i \uparrow\uparrow U$, whereas a
countercurrent propagating incident waves is denoted by $k_i
\downarrow\uparrow U$. The acronyms PEW and NEW pertain to
positive and negative energy waves, correspondingly.\\

\noindent\begin{tabular}{|l|c|c|c|} \hline
\multicolumn{4}{|c|}{\textbf{I. Subcritical
flow in the upstream and downstream domains}}\\
\hline %
\;${\bf k}_i$, ${\bf U}$ & {\bf Reflect. coeff.} & {\bf Transmiss. coeff.} & {\bf Peculiarity of a scattering} \\
\hline %
$k_i \uparrow\uparrow U$ & $R_\eta$ see Eq.~(\ref{TransCoef1})
 & $T_\eta$ see Eq.~(\ref{TransCoef1}) & Regular scattering \\
\hline %
$k_i \downarrow\uparrow U$ & $R_\eta$ see
Eq.~(\ref{TransCoef0.22}) & $T_\eta$ see Eq.~(\ref{TransCoef0.22}) & Regular scattering \\
\hline %
\multicolumn{4}{|c|}{\textbf{II. Subcritical flow in the upstream domain and supercritical in the}}\\
\multicolumn{4}{|c|}{\textbf{downstream domain. PEW and NEW appear downstream.}}\\
\hline%
\;${\bf k}_i$, ${\bf U}$ & {\bf Reflect. coeff.} & {\bf Transmiss. coeff.} & {\bf Peculiarity of a scattering} \\
\hline %
$k_i \uparrow\uparrow U$ & $R_{\eta}$ is undetermined, &
$T_{p}$ see Eq.~(\ref{Sub-superT1}) & Undefined problem statement, \\
 & according to \citep{ChurErStep-2017}, $R_{\eta} = 1$ & $T_{n}$ see Eq.~(\ref{Sub-superT2}) & according to \citep{ChurErStep-2017}, $T_p = -T_n = 1$  \\
\hline%
$k_i \downarrow\uparrow U$ & \multicolumn{3}{c|}{Impossible situation} \\
\hline \multicolumn{4}{|c|}{\textbf{III. Supercritical flow in the upstream and downstream domains}}\\
\hline%
\;${\bf k}_i$, ${\bf U}$ & {\bf Reflect. coeff.} & {\bf Transmiss. coeff.} & {\bf Peculiarity of a scattering} \\
\hline %
$k_i \uparrow\uparrow U$ & No reflected wave &
$T_{p}$ see Eq.~(\ref{Super-superT1}) & Incident wave can be PEW or NEW,\\
 &  & $T_{n}$ see Eq.~(\ref{Super-superT2}) & or both. See Eqs.~(\ref{Super-superT3}), (\ref{Super-superT5}).\\
\hline%
$k_i \downarrow\uparrow U$ & \multicolumn{3}{c|}{Impossible situation} \\
\hline \multicolumn{4}{|c|}{\textbf{IV. Supercritical flow in the upstream and}}\\
\multicolumn{4}{|c|}{\textbf{subcritical in the downstream domain}}\\
\hline%
\;${\bf k}_i$, ${\bf U}$ & {\bf Reflect. coeff.} & {\bf Transmiss. coeff.} & {\bf Peculiarity of a scattering} \\
\hline %
$k_i \uparrow\uparrow U$ & No reflected wave & $T_{\eta}$ provided that & Over-determined problem if \\
 &  & $A_n \sim A_p$, Eq.~(\ref{RelApAn}) & there is only one incident wave \\
\hline%
$k_i \downarrow\uparrow U$ & Formally $R_\eta = 1$ & Formally $T_\eta = 0$ & See Ref. \citep{ChurErStep-2017}\\
\hline
\end{tabular}\\

\bigskip

The problem studied can be further generalized for waves of
arbitrary length taking into account the effect of dispersion.
Similar works in this direction were published recently for
relatively smooth current variation in the canal with the
finite-length bottom obstacles \citep{RobMichPar-2016,
CoutWein-2016}. It is worthwhile to notice that in the dispersive
case for purely gravity waves there is always one wave of negative
energy for which the flow is supercritical. This negative energy
mode smoothly transforms into the dispersionless mode when the
flow increases. In such cases two other upstream propagating modes
disappear, and the dispersion relations reduces to one of
considered in this paper. It will be a challenge to compare the
theoretical results obtained in this paper with the numerical and
experimental data; this may be a matter of future study.

\begin{acknowledgments}
This work was initiated when one of the authors (Y.S.) was the
invited Visiting Professor at the Institut Pprime, Universit{\'e}
de Poitiers in August--October, 2016. Y.S. is very grateful to the
University and Region Poitou-Charentes for the invitation and
financial support during his visit. Y.S. also acknowledges the
funding of this study from the State task program in the sphere of
scientific activity of the Ministry of Education and Science of
the Russian Federation (Project No. 5.1246.2017/4.6), and G.R.
acknowledges the funding from the ANR Grant HARALAB No.
ANR-15-CE30-0017-04. The research of A.E. was supported by the
Australian Government Research Training Program Scholarship.

The authors are thankful to Florent Michel, Renaud Parentani,
Thomas Philbin, and Scott Robertson for useful discussions.
\end{acknowledgments}

\appendix
\section{Derivation of time-averaged wave-energy density for gravity
waves on a background flow}%
\label{appA}%

Here we present the derivation of the time averaged wave energy
density of traveling gravity surface wave on a background flow in
shallow water when there is no dispersion. In the linear
approximation on wave amplitude the depth integrated density of
wave energy (``pseudo-energy'' according to the terminology
suggested by \citet{McIntyre-1981}) can be defined as the
difference between the total energy density of water flow in the
presence of a wave and in the absence of a wave (we remind the
reader that in such approximation the wave energy density is
proportional to the squared wave amplitude):
\begin{equation}
\label{A1}%
E = \left\langle\left[\int\limits_{0}^\eta\rho g z\,dz +
\frac{\rho}{2}\int\limits_{-h}^\eta\left(U + u\right)^2\,dz
\right] - \frac{\rho}{2}\int\limits_{-h}^0 U^2\,dz, \right\rangle,
\end{equation}
where the angular brackets stand for the averaging over a period.
The first two terms in the square brackets represent the sum of
potential and total kinetic energies, whereas the negative terms
in the angular brackets represent the kinetic energy density of a
current per se. Removing the brackets and retaining only the
quadratic terms, we obtain (the linear terms disappear after the
averaging over time, whereas the cubic and higher-order terms are
omitted as they are beyond the accuracy in the linear
approximation):
$$
E = \left\langle \frac{\rho g}{2}\eta^2 +
\frac{\rho}{2}\int\limits_{-h}^0\left(U + u\right)^2\,dz +
\frac{\rho}{2}\int\limits_{-h}^\eta 2Uu\,dz -
\frac{\rho}{2}\int\limits_{-h}^0U^2\,dz \right\rangle
$$
\begin{equation}
\label{A2}%
{} = \frac{\rho g}{2}\left\langle \eta^2 \right\rangle +
\frac{\rho h}{2}\left\langle u^2 \right\rangle + \left\langle \rho
U \int\limits_{-h}^0 u\,dz + \rho U \int\limits_{0}^\eta u\,dz
\right\rangle.
\end{equation}

In the last angular brackets the first integral disappears after
averaging over a period of sinusoidal wave, and the last integral
for perturbations of infinitesimal amplitude can be presented in
accordance with the ``mean value theorem for integrals'' as the
product $u\eta$. Then, the energy density reads:
\begin{equation}
\label{A3}%
E = \frac{\rho g}{2}\left\langle \eta^2 \right\rangle + \frac{\rho
h}{2}\left\langle u^2 \right\rangle + \rho U \left\langle u\eta
\right\rangle.
\end{equation}

Eliminating $u$ with the help of Eq. (\ref{Rel-u&eta}), we obtain
for the downstream and upstream propagating waves
\begin{equation}
\label{A4}%
E = \left(\frac{\rho g}{2} + \frac{\rho}{2h}c_0^2 \pm \frac{\rho
Uc_0}{h} \right)\left\langle \eta^2 \right\rangle = \rho g\left(1
\pm \mathrm{Fr}\right)\left\langle \eta^2 \right\rangle,
\end{equation}
where sign plus pertains to the downstream propagating wave and
sign minus -- to the upstream propagating wave.

Thus, we see that the wave energy density is negative when
$\mathrm{Fr} > 1$, i.e., when a wave propagates against the
current. In the meantime, the dispersion relation in a shallow
water can be presented as $\omega = c_0|k| + {\bf U}\cdot{\bf k}$,
so that for the cocurrent propagating wave with $k > 0$ we have
$\omega = (c_0 + U)k = c_0k(1 + \mathrm{Fr})$, whereas for the
countercurrent propagating waves with $k < 0$ we have $\omega =
(c_0 - U)|k| = c_0k(\mathrm{Fr} - 1)$ (see Eq. (\ref{DispRelRef})
and explanation of Fig. \ref{f04}). Then the group velocity $V_g =
d\omega/dk = c_0(\mathrm{Fr} - 1)$ is positive if $\mathrm{Fr} >
1$ and negative if $\mathrm{Fr} < 1$. Hence, the wave energy flux
for the negative energy waves in the supercritical case with
$\mathrm{Fr} > 1$ is $W \equiv EV_g < 0$ and directed against the
group velocity.

Notice in the conclusion that the relationship between the wave
energy and frequency follows directly from the conservation of
wave action density $N$ (see Ref. \cite{MaiRusStep-2016A} and
references therein):
\begin{equation}
\label{A5}%
N = \frac{E}{\omega - {\bf U}\cdot{\bf k}} = \frac{E_0}{\omega},
\end{equation}
where $E$ is the density of wave energy in the immovable
coordinate frame (\ref{A4}) where the water flows with the
constant speed ${\bf U}$, and $E_0$ and $\omega = c_0|{\bf k}|$
are the density of wave energy and frequency in the coordinate
frame moving with the water.

\bibliography{ChERS}

\end{document}